\documentclass[10pt,letterpaper]{article}
\usepackage{opex3}
\usepackage{amsmath}

\begin{document}

\title{Relation between surface solitons and bulk solitons in nonlocal nonlinear media}

\author{Zhenjun Yang,$^{1,2}$ Xuekai Ma,$^1$ Daquan Lu,$^1$
Yizhou Zheng,$^1$  Xinghui Gao,$^1$ and Wei Hu,$^{1,*}$ }

\address{$^1$Laboratory of Photonic Information Technology, South
China Normal University, \\ Guangzhou 510631, P. R. China}
\address{$^2$College of Physics Science and Information Engineering,
Hebei Normal University, \\ Shijiazhuang 050016, P. R. China}

\email{huwei@scnu.edu.cn} 



\begin{abstract}
We find that a surface soliton in nonlocal nonlinear media can be regarded as a
half of a bulk soliton with an antisymmetric amplitude distribution. The
analytical solutions for the surface solitons and breathers in strongly
nonlocal media are obtained, and the critical power and breather period are
gotten analytically and confirmed by numerical simulations. In addition, the
oscillating propagation of  nonlocal surface solitons launched away from the
stationary position is considered as the interaction between the soliton and
its out-of-phase image beam. Its trajectory and oscillating period obtained by
our model are in good agreement with the numerical simulations.
\end{abstract}



\section{Introduction}
The nonlocality of the nonlinear response exists in many real
physical systems, such as photorefractive crystals
\cite{Mitchell1998-PRL}, nematic liquid crystals
\cite{Peccianti2002-OL,Conti2003-PRL,Conti2004-PRL,Hu2006-APL}, lead glasses
\cite{Rotschild2005-PRL,Rotschild2006-OL}, atomic vapors
\cite{Skupin2007-PRL}, Bose-Einstein condensates
\cite{Pedri2005-PRL,Tikhonenkov2008-PRL} etc. In nonlocal nonlinear
media, various soliton solutions have been predicted theoretically,
such as vortex solitons \cite{Kartashov2007-OE1,Lu2009-PRA},
multi-pole solitons
\cite{Rotschild2006-OL,Xu2005-OL,Dong2010-PRA,Ouyang2006-PRE},
Laguerre-Gaussian and Hermite-Gaussian solitons
\cite{Buccoliero2007-PRL,Deng2007-JOSAB,Deng2008-JOA}, Ince-Gaussian
solitons \cite{Deng2007-OL}, and some of these have been observed
experimentally \cite{Conti2004-PRL,Hu2006-APL,Rotschild2005-PRL,Rotschild2006-OL,
Izdebskaya2011-OL}. For nonlocal solitons,
there are many interesting properties, for instance, the large phase
shift \cite{Guo2004-PRE}, attraction between two bright out-of-phase
solitons \cite{Hu2006-APL,Rasmussen2005-PRE}, attraction between two
dark solitons \cite{Dreischuh2006-PRL,Nikolov2004-OL}, self-induced fractional Fourier
transform \cite{Lu2008-PRA}, etc.

Recently, the nonlocal surface solitons have been investigated
numerically and experimentally
\cite{Alfassi2007-PRL,Kartashov2009-OL,Ye2008-PRA,Kartashov2007-OE2,Alfassi2009-PRA,Kartashov2008-OL}.
The nonlocal surface solitons occurring at the interface between a
dielectric medium and a nonlocal material exhibit unique properties.
Nonlocal multipole surface solitons, vortices, and bound states of
vortex solitons, incoherent surface solitons and ring surface waves
have been also studied recently
\cite{Kartashov2009-OL,Ye2008-PRA,Kartashov2007-OE2,Alfassi2009-PRA}.
But the properties of nonlocal surface solitons mentioned above are
all discussed by numerical simulations. In order to get a good
understanding of the properties of nonlocal surface solitons, it is
essential to present an analytic solution, even an approximate one.

In this paper, based on the propagation equations governing the nonlocal
surface waves and assuming that all energy of the surface soliton resides in
nonlocal media, we find a surface soliton in nonlocal nonlinear media can be
regarded as a half of a bulk soliton with an antisymmetric amplitude
distribution. The evolution regularities for the nonlocal surface waves are
discussed both analytically and numerically. The analytical solutions for the
surface solitons and breathers in strongly nonlocal media are obtained, and the
critical power and breather period are gotten analytically and confirmed by
numerical simulations. In addition, the oscillating propagation of nonlocal
surface solitons launched away from the stationary position is considered as
the interaction between the soliton and its out-of-phase image beam. Its
trajectory and oscillating period obtained by our model are in good agreement
with the numerical simulations.

\section{Relation between surface solitons and bulk solitons}\label{theory}

\begin{figure}[htbp]
\centering\includegraphics[width=7cm]{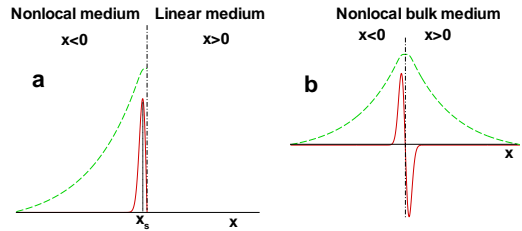} \caption{Sketches of a
nonlocal surface soliton (a) and a nonlocal antisymmetric bulk soliton (b).
Solid and dashed lines represent the distributions of the amplitude and the
nonlinear refraction index, respectively.}\label{SetSketch}
\end{figure}

We consider a (1+1)-D model of an optical surface wave with an envelope $q$
propagating in $z$ direction near an interface between a nonlocal nonlinear
medium and a linear medium [see Fig. \ref{SetSketch}(a)]. The propagation of
surface waves is governed by the dimensionless nonlocal nonlinear
Schr\"{o}dinger equation (NNLSE),  i.e. \\
(i) in nonlocal nonlinear media, $x \leq0$,
\begin{eqnarray}
i\frac{\partial q}{\partial
z}+\frac{1}{2}\frac{\partial^2q}{\partial x^2}+\Delta nq &=&0,\label{Nonlocal-Propagation} \\
\Delta n-w_m^2\frac{\partial^2\Delta n}{\partial
x^2}&=&|q|^2;\label{Nonlocal-Response}
\end{eqnarray}
(ii) in linear media, $x>0$,
\begin{equation}
i\frac{\partial q}{\partial
z}+\frac{1}{2}\frac{\partial^2q}{\partial
x^2}-qn_d=0,\label{Linear-Propagation}
\end{equation}
where $x, z$ denote, respectively, the normalized transversal and
longitudinal coordinates. $\Delta n$ is the nonlinear perturbation
of the refractive index, and $n_d$ denotes the normalized difference between
the unperturbed refractive index of the nonlocal medium and the
refractive index of the less optically dense linear medium. $w_m$
represents the characteristic length of the nonlocal material
response, and $\alpha=w_m/w_0$ is the degree of nonlocality, where
$w_0$ is the beam width. The optical intensity is normalized as $I=|q|^2$. In fact, Eqs. (\ref{Nonlocal-Propagation})
and (\ref{Nonlocal-Response}) denote a NNLSE with an
exponential-decay type nonlocal response. The exponential-decay type
nonlocal response exists in many real physical systems, for
instance, all diffusion-type nonlinearity
\cite{Krolikowski2001-PRE,Ghofraniha2007-PRL}, orientational-type
nonlinearity \cite{Peccianti2002-OL,Rasmussen2005-PRE}, and the
general quadratic nonlinearity describing parametric interaction
\cite{Nikolov2003-PRE,Larsen2006-PRE}. Moreover when
$w_m\rightarrow\infty$, Eqs. (\ref{Nonlocal-Propagation}) and
(\ref{Nonlocal-Response}) can be transformed to the forms describing
the thermal nonlocal media (for example, lead glasses\cite{Rotschild2005-PRL,Alfassi2007-PRL}).

The boundary conditions for the surface soliton are
$q(x\rightarrow-\infty)=0$, $q(+0)=q(-0)$, $\Delta n(x\rightarrow
-\infty)=0$, and $\partial\Delta n/\partial x|_{x=0}=0$
\cite{Alfassi2007-PRL,Kartashov2009-OL,Ye2008-PRA}. References
\cite{Alfassi2007-PRL,Kartashov2009-OL,Ye2008-PRA} indicate if the
index difference $n_d$ is big enough, namely $n_d\gg1$, which is
easily satisfied in the actual physical system, the optical energy
is almost totally confined in the nonlocal medium. Therefore one can
approximately get the relation $q(-0)=q(+0)=0$ and $q(x>0)=0$. Under
this approximation, the propagation of surface waves can be solved
totally only based on Eqs. (\ref{Nonlocal-Propagation}) and
(\ref{Nonlocal-Response}) with the boundary conditions ($x\leq0$),
\begin{subequations}\label{SurfaceConditions}
\begin{align}
q(x\rightarrow-\infty)=q(0)=0, \\
\left.\frac{\partial\Delta n}{\partial x}\right|_{x=0}=0.
\end{align}
\end{subequations}

We find the solution of a surface soliton under the approximation
$q(0)=0$ is identical with the half part of an antisymmetric soliton
in a bulk medium as shown in Fig. \ref{SetSketch}(b). For the bulk
soliton solutions with antisymmetric amplitude distribution which is
also governed by Eqs. (\ref{Nonlocal-Propagation}) and
(\ref{Nonlocal-Response}), one can easily obtain the following
relations.
\begin{subequations}\label{BulkConditions}
\begin{align}
q(-0)=q(+0)=0,  \\
q(x\rightarrow\pm\infty)=0,  \\
\left.\frac{\partial\Delta n}{\partial x}\right|_{x=0}=0.
\end{align}
\end{subequations}
Comparing Eq. (\ref{SurfaceConditions}) with Eq.
(\ref{BulkConditions}), it can be found that the conditions that the
surface solitons satisfy are the same as those of the left half
($x\leq0$) of the antisymmetric bulk solitons. Therefore, the
surface soliton can be regarded as a half of the bulk soliton, and
all the results of antisymmetric bulk soliton can be transfered to
the surface soliton. On the other hand, the right half of the bulk
soliton can be regarded as an image beam of the surface soliton
\cite{Shou2009-OL}. The interaction between the surface soliton and
the interface can be regarded as the interaction between the soliton
and its image beam in bulk medium.

In nonlocal bulk media, the soliton solutions can have the
antisymmetric amplitude distribution
\cite{Xu2005-OL,Dong2010-PRA,Ouyang2006-PRE}. Some authors have
demonstrated that Hermite-Gaussian function can be applied to
describe this type soliton solution in nonlinear media with several
different nonlocal responses, especially for the strongly nonlocal
case \cite{Ouyang2006-PRE,Buccoliero2007-PRL,Deng2007-JOSAB}. In
addition, it has been discovered that in a nonlinear material with a
finite-range nonlocal or a very long-range nonlocal response, the
maximal number of peaks in stable multipole bulk solitons is four,
and all higher-order soliton bound states are unstable
\cite{Xu2005-OL,Dong2010-PRA}. Therefore, surface solitons only with
less than three poles can be stable in the case of thermally
nonlocal interface, which is firstly addressed by Kartashov {\it et
al}\cite{Kartashov2009-OL}.

In the following, we give some comparisons between the bulk solitons
and the surface solitons to illustrate our conclusions, and some
analytical results are given for the nonlocal surface solitons.

\section{Surface solitons and breathers}\label{SolitonandBreather}

In strongly nonlocal nonlinear media, there exist Hermite-Gaussian solitons or
breathers\cite{Ouyang2006-PRE,Buccoliero2007-PRL,Deng2007-JOSAB}. Because the
first-order Hermite-Gaussian beam is antisymmetric about the beam center, it
can be used to describe the nonlocal surface soliton. In bulk nonlocal media,
the first-order Hermite-Gaussian trial beams can be expressed as
\begin{equation}\label{TrialSolution}
q(x,z)=a(z)x\exp\left[-\frac{x^2}{w^2(z)}+ic(z)x^2\right]e^{i\theta(z)},
\end{equation}
where $a(z), w(z), c(z), \theta(z)$ represent the amplitude, beam
width, phase-front curvature and phase of the beams, respectively.
The input power is $P_0=\int |q|^2dx$.

In the following, Eqs. (\ref{Nonlocal-Propagation})
and (\ref{Nonlocal-Response}) are used for the equivalent bulk
case ($-\infty< x <\infty$). Therefore Eqs.(\ref{Nonlocal-Propagation}) and (\ref{Nonlocal-Response})
can be restated as an Euler-Lagrange equation corresponding to a
variational problem
\cite{Buccoliero2007-PRL,Anderson1983-PRA,Cao2008-OC}
\begin{equation}\label{DetaL}
\delta\int_{0}^{+\infty}[L]dz=0,
\end{equation}
where
\begin{equation}\label{[L]}
[L]=\int_{-\infty}^{+\infty}L(q,q^*,\frac{\partial q}{\partial
x},\frac{\partial q^*}{\partial x},\frac{\partial q}{\partial
z},\frac{\partial q^*}{\partial z})dx.
\end{equation}
The Lagrangian density $L$ is given by
\begin{eqnarray}\label{Lagrange}
L=\frac{i}{2}\left(q^*\frac{\partial q}{\partial z}-q\frac{\partial
q^*}{\partial z}\right)-\frac{1}{2}\left|\frac{\partial q}{\partial
x}\right|^2+\frac{1}{2}\left|q\right|^2\int^{+\infty}_{-\infty}R(x-x')\left|q(x')\right|^2dx',
\end{eqnarray}
where $R(x)=(1/2w_m)\exp(-|x|/w_m)$. Because the complexity of
integration produced by $R(x)$, the expression of $[L]$ is too
difficult to obtain. But for the strongly nonlocal case, i.e.
$\alpha=w_m/w_0\gg 1$, $R(x)$ can be expanded as
\begin{equation}\label{R-2}
R(x)\approx\frac{1}{2w_m}\left(1-\frac{|x|}{w_m}+\frac{x^2}{2w^2_m}\right).
\end{equation}

Combining Eq. (\ref{TrialSolution}) and Eq. (\ref{[L]}), $[L]$ can
be obtained
\begin{eqnarray}\label{[L]-expression}
[L]&=&-\frac{3}{8}\sqrt{\frac{\pi}{2}}
a^2w\left(1+c^2w^4\right)-\frac{1}{16}\sqrt{\frac{\pi}{2}}a^2w^3\left(3w^2
\frac{dc}{dz}+4\frac{d\theta}{dz}\right)\nonumber\\
&&+\frac{a^4w^6}{512w_m^3}\left(4\pi w_m^2-7\sqrt{\pi}w_mw+3\pi
w^2\right).
\end{eqnarray}
Following the common process of the variational approach, one can
get four differential equations,
\begin{subequations}\label{Euler}
\begin{align}
&2w\frac{da}{dz}+3a\frac{dw}{dz}=0,\label{Euler-1} \\
&2w\frac{da}{dz}+a\left(5\frac{dw}{dz}-4cw\right)=0,\label{Euler-2} \\
&\frac{a^2w^5}{w_m^3}\left(4\pi w_m^2-7\sqrt{\pi}w_mw+3\pi w^2\right)\nonumber\\
&-48\sqrt{2\pi}\left(1+c^2w^4\right)
-8\sqrt{2\pi}w^2\left(3w^2\frac{dc}{dz}+4\frac{d\theta}{dz}\right)=0,\label{Euler-3} \\
&480\sqrt{2}w_m^3c^2w^4+a^2w^5\left(-24\sqrt{\pi}w_m^2+49w_mw-24
\sqrt{\pi}w^2\right)\nonumber\\
&+48\sqrt{2}w_m^3\left(4w^2
\frac{d\theta}{dz}+5w^4\frac{dc}{dz}+2\right)=0.\label{Euler-4}
\end{align}
\end{subequations}
Based on the above four equations, we can investigate the
propagation of the trial beams. In the following, we divide the
problem into two cases to discuss, i.e. solitons and breathers.

\subsection{surface solitons}

For the case of solitons, in Eq. (\ref{TrialSolution}), $a$ and $w$ reduce to
constants, and $c=0$, $\theta(z)=\beta z$, where $\beta$ is propagation
constant. The expression of beams can be rewritten in a simple form
\begin{equation}\label{BulkSoliton}
q(x,z)=a_0x\exp\left(-\frac{x^2}{w^2_0}\right)e^{i\beta z}.
\end{equation}
Based on Eq. (\ref{Euler}), we can get the input power, namely the
critical power of bulk solitons
\begin{equation}\label{SolitonPower}
P_c=\frac{48\sqrt{\pi}w_m^3}{w_0^3(7w_m-6\sqrt{\pi}w_0)},
\end{equation}
and the propagation constant
\begin{equation}\label{beta}
\beta=\frac{105w_mw_0-6\sqrt{\pi}(8w_m^2+9w_0^2)}
{2w_0^3(6\sqrt{\pi}w_0-7w_m)}.
\end{equation}

\begin{figure}[htbp]
\centering
\includegraphics[width=7cm]{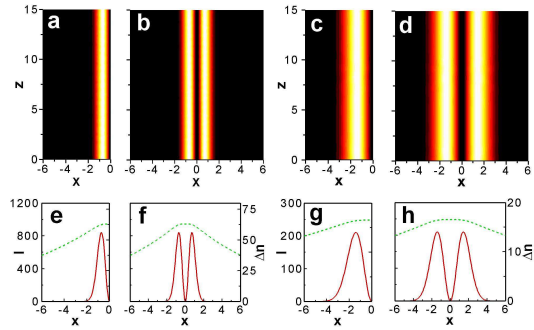}
\caption{Comparison between the fundamental surface solitons and the
first-order Hermite-Gaussian bulk solitons with the degree of nonlocality
$\alpha=10$.  (a) and (c) are simulated intensity distribution for surface
solitons during propagation with $w_0=1.0$ and $2.0$, respectively; (b) and (d)
are simulated intensity distribution for bulk solitons with $w_0=1.0$ and
$2.0$, respectively. (e)-(h) are, respectively, the transversal intensity
distributions (solid line) and the refractive index distributions (dashed line)
corresponding to row 1.}\label{SolitonPropagation}
\end{figure}

\begin{figure}[htbp]
\centering
\includegraphics[width=7cm]{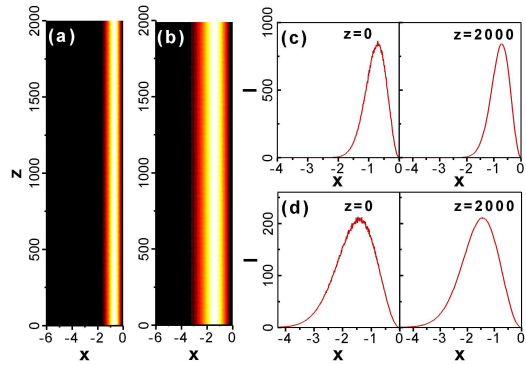}
\caption{Simulated propagation of surface solitons
in the presence of 3\% white noise at a fixed nonlocal degree
$\alpha=10$. (a) and (b) give the propagation of surface solitons
with different soliton widths $w_0=1.0$ and $2.0$, respectively. (c)
and (d) give the incident intensity distribution (left) and the
output intensity distribution at $z=2000$ (right) corresponding to
(a) and (b), respectively.}\label{Stability}
\end{figure}

According to our conclusion in Sec. \ref{theory}, the surface
soliton can be expressed as
\begin{equation}\label{SurfaceSoliton}
q(x,z)=\begin{cases} a_0x
\exp\left(-\frac{x^2}{w^2_0}\right)e^{i\beta z} &\text{for}~~~x\leq0,\\
0 &\text{for}~~~x>0.
\end{cases}
\end{equation}
Apparently, the critical power of surface soltions is a half of bulk
solitons, namely $P_s=P_c/2$. When the optical beam propagates as a
surface soliton, we define the position where the beam has its
maximal intensity as the soliton position, i.e. the distance between
the maximal intensity and the interface. According to Eq.
(\ref{SurfaceSoliton}), the soliton position $x_s$ can be obtained
as
\begin{equation}\label{SolitonPosition}
x_s=-\frac{\sqrt{2}w_0}{2}.
\end{equation}

Based on Eqs. (\ref{Nonlocal-Propagation})-(\ref{Linear-Propagation}), some
numerical simulations of the propagations of surface and bulk solitons are
carried out by using Eqs. (\ref{BulkSoliton}) and (\ref{SurfaceSoliton}) as
incident profiles. Figure \ref{SolitonPropagation} shows the analytical and simulated results
are in good agreement with each other in strongly nonlocal media.
In fig. \ref{SolitonPropagation}, $I=|q|^2$ is the optical intensity.
To confirm the stability, we simulate the propagation of surface
solitons in the presence of $3\%$ white noise (see fig.
\ref{Stability}). Figure \ref{Stability} shows the propagation is
stable.

\begin{figure}[htbp]
\centering
\includegraphics[width=7cm]{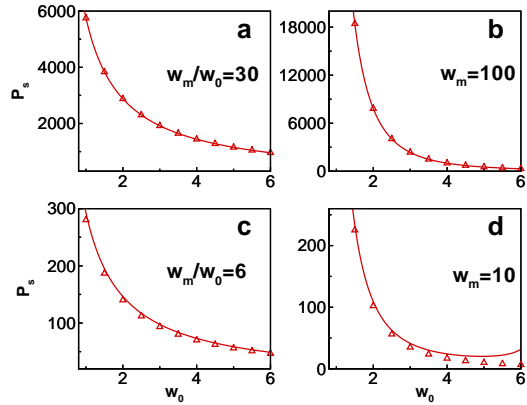}
\caption{Variations of the surface soliton power with the soliton width under
different conditions. (a) and (c) are the results with a fixed degree of
nonlocality $\alpha=30, 6$, respectively. (b) and (d) are the results with a
fixed width of the nonlocal response $w_m=100, 10$, respectively. Solid lines
represent the analytical results based on Eq. (\ref{SolitonPower}); the
triangles represent the numerical results.}\label{PcVsW0}
\end{figure}

\begin{figure}[htbp]
\centering
\includegraphics[width=7cm]{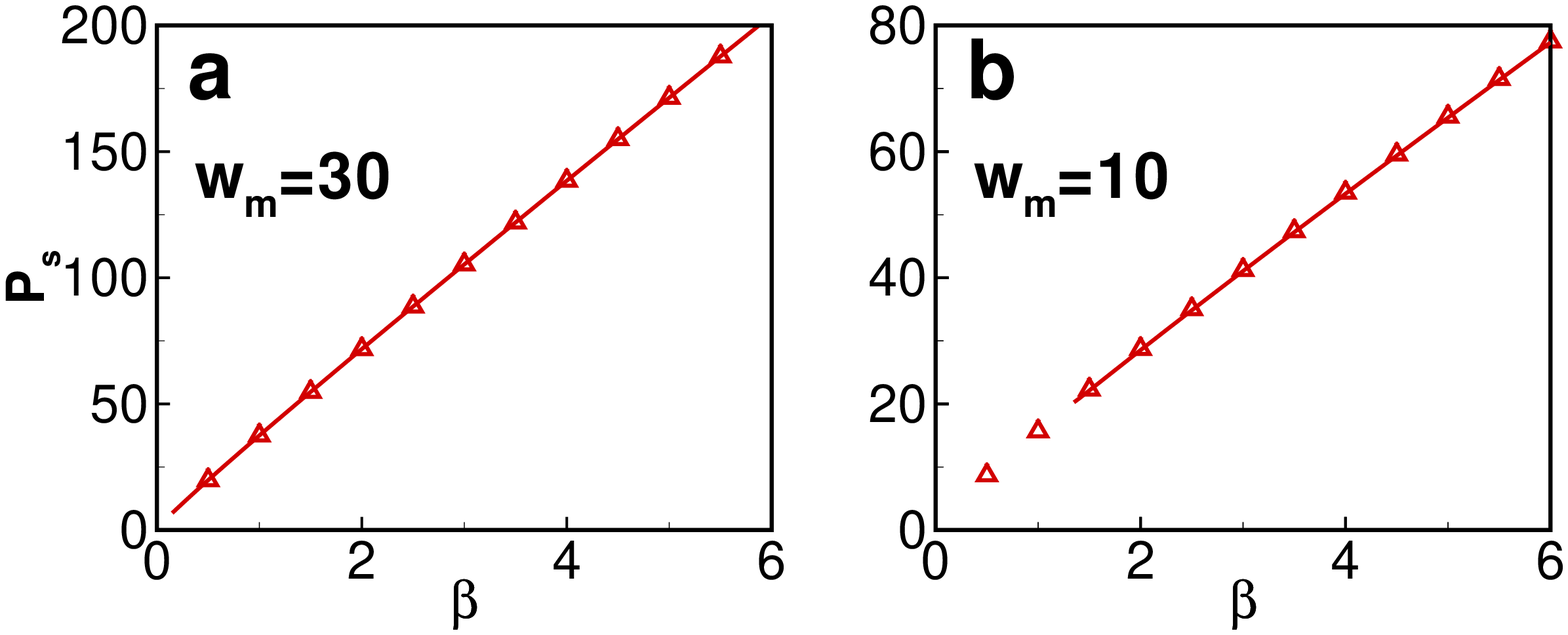}
\caption{Variations of the surface soliton power with the propagation constant
for (a) $w_m=30$ and (b) $w_m=10$. Solid lines represent the analytical results
based on Eqs. (\ref{SolitonPower}) and (\ref{beta}); the triangles represent
the numerical results.}\label{PcVsBeta}
\end{figure}

In general nonlocal media, the profiles of surface soliton are obtained by numerical iterative method based on Eqs.
(\ref{Nonlocal-Propagation})-(\ref{Linear-Propagation}), and the relations
between  critical powers and beam width are shown in Fig. \ref{PcVsW0}. For
Figs. \ref{PcVsW0}(a) and \ref{PcVsW0}(b) which belong to the strongly nonlocal
case, the analytical results are in excellent agreement with the numerical
results. As the degree of nonlocality decreases to $\alpha=6$ in Fig.
\ref{PcVsW0}(c), the analytical results are also accordant with the numerical
results approximately. For Fig. \ref{PcVsW0}(d), the analytical results begin
to deviate from the exact numerical ones with the beam width increasing. Figure
\ref{PcVsW0} confirms that the analytical solutions of surface solitons are
very valid for the strongly nonlocal case. The relations between critical
powers and propagation constants are shown in Fig. \ref{PcVsBeta}. It shows the
analytical results from Eqs. (\ref{SolitonPower}) and (\ref{beta}) are in good
agreement with the numerical results except for small $\beta$ which corresponds
to a weak nonlocality.

\begin{figure}[htbp]
\centering
\includegraphics[width=7cm]{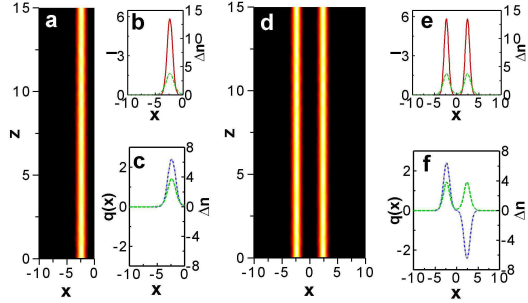}
\caption{Comparison between the surface solitons and the bulk solitons for the
weakly nonlocal case ($\alpha=0.2$). (a) and (d) are the simulated intensity
distributions for the surface soliton and the bulk soliton during propagation
with $w_0=1.0, w_m=0.5$, respectively; (b) and (e) are the transversal
intensity distributions (solid line) and the refractive index distributions
(dashed line) corresponding to (a) and (d), respectively. (c) and (f) are the
transversal amplitude distributions (dashed-dotted line) and the refractive
index distributions (dashed line) corresponding to (a) and (d),
respectively.}\label{WeakSolitonPropagation}
\end{figure}

For weakly nonlocal case, the analytical solution can not be
obtained because the analytical solution for multipole bulk solitons
can not be found in weakly nonlocal medium. However, in nonlocal
bulk media, the diploe solitons are always existent and stable in
the entire domain of their existence \cite{Xu2005-OL}. According to
the theory in Sec. \ref{theory}, the fundamental surface soliton
should be existent for the weakly nonlocal case. Following the
method in Ref. \cite{Xu2005-OL}, we seek the dipole bulk soliton
combining two out-of-phase fundamental solitons under weakly
nonlocality, as a result, the surface soliton can be also found at
the same time, as shown in Fig. \ref{WeakSolitonPropagation}.

\subsection{surface breathers}

Because the breathers exist in nonlocal bulk media, if the input
power is not equal to the soliton power, we can find the surface
breathers. According to Eq. (\ref{Euler}), we can get
\begin{equation}\label{w2}
\frac{d^2w}{dz^2}=\frac{2\sqrt{2}}{w^3}-\frac{7a^2w^3}{96\sqrt{2}w_m^2}+
\frac{\sqrt{\pi}a^2w^4}{16w_m^3}.
\end{equation}
Above equation is equivalent to Newton's second law in classical
mechanics for the motion of an one-dimensional particle with the
equivalent mass 1, thus one can get the equivalent potential
\begin{equation}\label{potential}
V(w)=\frac{\sqrt{2}}{w^2}-\frac{a^2\sqrt{\pi}w^5}{80w_m^3}+\frac{7a^2w^4}{384w_m^2}+c_0,
\end{equation}
where $c_0$ is a constant. Through expanding the equivalent potential $V(w)$
into the second order at the balance position, the motion of the particle is
approximated to a harmonic oscillation, and we can obtain the approximate
expression of the beam width
\begin{equation}\label{BreatherWidth}
w(z)=\frac{\sqrt{2}}{2}\left[g-\sqrt{b}+(\sqrt{2}w_0-g+\sqrt{b})
\cos\left(\sqrt{\frac{12}{(g-\sqrt{b})^4}
+\frac{1}{3w_0p}}z\right)\right],
\end{equation}
where
\begin{eqnarray}\label{BreatherParameters}
g&=&\frac{7w_m}{12\sqrt{2\pi}}+\sqrt{t+\frac{49w_m^2}{288\pi}},\nonumber\\
b&=&g^2-2t-2\sqrt{t^2+3w_0p},\nonumber\\
t&=&(h+\sqrt{p^3+h^2})^{\frac{1}{3}}
     +(h-\sqrt{p^3+h^2})^{\frac{1}{3}},\nonumber\\
h&=&-\frac{49w_0w_m^2}{192\pi},\nonumber\\
p&=&\frac{w_0^2P_c(6\sqrt{\pi}w_0-7w_m)}{18\sqrt{\pi}P_{0}}.\nonumber
\end{eqnarray}

\begin{figure}[htbp]
\centering
\includegraphics[width=7cm]{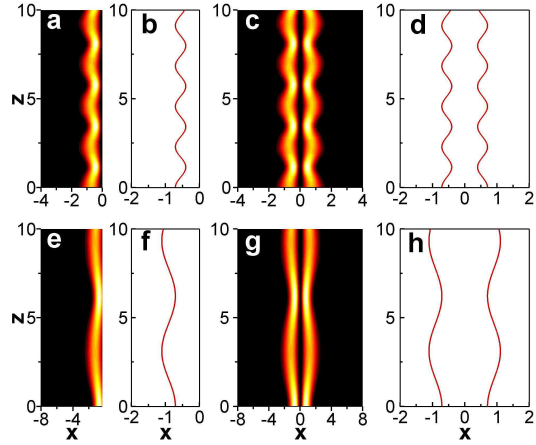}
\caption{Comparison between the surface breathers and the bulk breathers with
$\alpha=10$. (a) and (c) are, respectively, the simulated propagations of the
surface breather and the bulk breather with input power being twice soliton
power. (b) and (d) give the analytical trajectories of the maximal intensity
corresponding to (a) and (c), respectively. Row 2 is the same as row 1 except
that the input power is a half of soliton power.}\label{BreatherPropagation}
\end{figure}

According to our theory, the oscillation period of surface breathers
is the same as that of bulk breathers. Then, we can obtain the
oscillation period of the surface breather,
\begin{equation}\label{BreatherPeroid}
\Delta z=2\pi\left[\frac{12}{(g-\sqrt{b})^4}
-\frac{6\sqrt{\pi}P_{s0}}{w_0^3P_s(7w_m-6\sqrt{\pi}w_0)}\right]^{-\frac{1}{2}},
\end{equation}
where $P_{s0}$ is the input power of the surface breathers. At the same time,
the beam trajectory can be obtained and given by $x_b=-\sqrt{2}w(z)/2$.  Figure
\ref{BreatherPropagation} gives the comparison between the surface breathers
and the bulk breathers for $\alpha=10$. The analytical trajectories are in good
agreement with the simulated results.

\section{Surface solitons launched away from the stationary position}

One can learn from Ref. \cite{Alfassi2007-PRL} that if the beam launched away
from the soliton position, the beam will propagate oscillating about the
stationary position. The oscillation induced by the interaction between the
soliton and the interface is periodic, and the beam never converges to a
straight line trajectory. According to our theory and introducing an
out-of-phase image beam \cite{Shou2009-OL}, the interaction between soliton and
interface can be regarded as the interaction between the soliton and its image
beam in nonlocal bulk media. Then the oscillating trajectory of surface
solitons can be obtained from the trajectories of two out-of-phase solitons
interacting in nonlocal bulk media.

For the strongly nonlocal case, the fields of two out-of-phase
Gaussian solitons propagating in bulk media can be expressed as
\begin{equation}\label{TwoGaussInitial}
q_{\pm}(x,z)=\pm a_0\exp\left[-\frac{[x\pm x_c(z)]^2}{w_0^2}\pm
iu(z)x\right]e^{i\theta (z)},
\end{equation}
where $\theta(z)$ is the phase of whole beams, $\pm x_c(z)$ are the
mass centers of Gaussian solitons and $x_c(0)=x_{c0}$. $u(z)$ is the
tilting wavefront induced by the attraction between two solitons,
and we consider the normally incident case, i.e. $u(0)=0$. The
amplitude
$a_0^2=32\sqrt{2}w_m^5/w_0^4[4w_m(w_0^2+2w_m^2)-\sqrt{\pi}w_0(w_0^2+4w_m^2)]$
is normalized to guarantee the incident power of each soliton is
equal to the critical power, so we can assume that the beam width
does not change during propagation.

\begin{figure}[htbp]
\centering
\includegraphics[width=7cm]{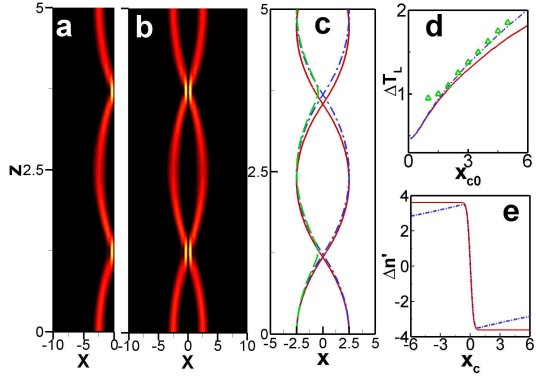}
\caption{(a) Simulated propagations of a Gaussian surface wave launched $2.5w_0$
away from the surface. (b) Simulated propagations of two out-of-phase Gaussian
solitons with a separated distance $5.0w_0$. (c) The trajectories of surface
wave and two Gaussian solitons. (d) The oscillation period of surface wave. (e)
Distribution of $\Delta n'$. The red solid line and the blue dashed-dotted line
are the results based on Eqs. (\ref{RayEquation}) and
(\ref{RayEquationApproximation}), respectively. The green dashed line and green
triangles are the simulated results directly based on Eqs.
(\ref{Nonlocal-Propagation})-(\ref{Linear-Propagation}). All cases are with the
parameters $w_m=50.0, w_0=1.0$.}\label{TwoGauss}
\end{figure}

If the two solitons are separated over $3w_0$ each other, their fields almost
do not overlap. However because of strong nonlocality, there still exists
attractive force between them. The attractive force exerted on one soliton is
produced by the other soliton, specifically, by the nonlinear refractive index
change induced by the other soliton. Therefore, the trajectory of one soliton
(i.e. the left one $q_+$) can be determined by the light ray equation,
\begin{equation} \label{RayEq}
\frac{d^2x_c}{dz^2}=\left.\frac{d(\Delta
n_{-})}{dx}\right|_{x=-x_c}\equiv \Delta n',
\end{equation}
where the refractive index change $\Delta n_{-}$ is only induced by
the right soliton $q_{-}$, as
\begin{eqnarray}\label{DeltaN}
\Delta n_{-}=\int_{-\infty}^{+\infty} R(x-x')|q_{-}(x',z)|^2dx'.
\end{eqnarray}
Using the nonlinear response function $R(x)=(1/2w_m)\exp(-|x|/w_m)$ and Eqs.
(\ref{TwoGaussInitial})-(\ref{DeltaN}),  one can obtain
\begin{eqnarray}\label{RayEquation}
\frac{d^2x_c}{dz^2}&=&\frac{\sqrt{\pi}a^2w_0}{4\sqrt{2}w_m^2}
\exp\left(\frac{w_0^2-16w_mx_c}{8w_m^2}\right)\nonumber\\
&&\times\left[\exp\left(\frac{4x_c}{w_m}\right)
\text{erfc}\left(\frac{w_0^2+8 w_m
x_c}{2\sqrt{2}w_0w_m}\right)+\text{erf}\left(\frac{w_0^2-8w_mx_c}{2
\sqrt{2}w_0w_m}\right)-1\right],
\end{eqnarray}
where $\text{erf}(\cdot)$ and $\text{erfc}(\cdot)$ denote the error and
complementary error functions, respectively. Taking the conditions $w_m\gg w_0,
x_{c0}$, we can get a simply approximate expression
\begin{eqnarray}\label{RayEquationApproximation}
\frac{d^2x_c}{dz^2}\simeq-\frac{a^2\sqrt{\pi}w_0}
{2\sqrt{2}w_m^2}\text{erf}\left(\frac{4 x_c}{\sqrt{2}w_0}\right),
\end{eqnarray}
$\Delta n'$ shown in Fig. \ref{TwoGauss}(e) represents the attractive force
between two solitons. Due to the complicated form of $\Delta n'$, Eqs.
(\ref{RayEquation}) and (\ref{RayEquationApproximation}) are solved numerically
using Runge-Kutta methods. The soliton trajectories and the oscillation period
are shown in Figs. \ref{TwoGauss}(c) and \ref{TwoGauss}(d). The trajectory of
the surface soliton corresponds to the left half of trajectories of two
solitons in bulk media, i.e. $x_{sc}=-|x_c|$.

The numerical simulations  based on Eqs. (\ref{Nonlocal-Propagation}) -
(\ref{Linear-Propagation}) for surface soliton launched away from stationary
position and for two bulk solitons are also shown in Figs. \ref{TwoGauss}(a)
and \ref{TwoGauss}(b).  The trajectories and the oscillation periods gotten
from numerical simulations are compared with that from  Eqs.
(\ref{RayEquation}) and (\ref{RayEquationApproximation}) in Figs.
\ref{TwoGauss}(c) and \ref{TwoGauss}(d). These results are approximately
coincident with each other. The main reason of disagreement is that Eq.
(\ref{RayEq}) is valid when two beam fields do not overlap. When $x_{c0}$ is
small, the fields of two beams overlap and we can not distinguish $\Delta
n_{+}$ and $\Delta n_{-}$. Then Eq. (\ref{RayEq}) is invalid, and the simulated
results disagree with that obtained by solving Eqs.(\ref{RayEquation})  and
(\ref{RayEquationApproximation}). When $x_{c0}$ is large and the nonlocality is
strong enough, the simulated results are in good agreement with the results by
numerically solving Eqs. (\ref{RayEquation}) and
(\ref{RayEquationApproximation}).

\section{Conclusion}

In conclusion, we have studied the nonlocal surface waves numerically and
analytically. We find a surface soliton in nonlocal nonlinear media can be
regarded as a half of a bulk soliton with an antisymmetric amplitude
distribution. By applying the variational method and taking the first-order
Hermite-Gaussian beam as an example, the analytical solutions for the surface
solitons and breathers in strongly nonlocal media are obtained, and the
critical power and breather period are gotten analytically and confirmed by
numerical simulations. In addition, the oscillating propagation of  nonlocal
surface soliton launched away from the stationary position is considered as the
interaction between the soliton and its out-of-phase image beam. We have
discussed the oscillation period and the beam trajectory. Its trajectory and
oscillating period obtained by our model are in good agreement with the
numerical simulations.

\section*{Acknowledgments}
This research was supported by the National Natural Science
Foundation of China (Grant Nos. 10804033 and 10674050), the Program
for Innovative Research Team of Higher Education in Guangdong (Grant
No. 06CXTD005), the Specialized Research Fund for the Doctoral
Program of Higher Education (Grant No. 200805740002), and the
Natural Science Foundation of Hebei Province (Grant No.
F2009000321).

\end{document}